\documentclass{elsart}
\usepackage{natbib}
\usepackage{graphicx}
\newcommand{\rgg}{($\gamma$,$\gamma'$)}
\newcommand{\rgn}{($\gamma$,${\lowercase{\rm{n}}}$)}
\newcommand{\rgp}{($\gamma$,${\lowercase{\rm{p}}}$)}
\newcommand{\rng}{(n,$\gamma$)}
\newcommand{\rpg}{(p,$\gamma$)}
\newcommand{\rga}{($\gamma$,$\alpha$)}

\begin{document}
\runauthor{P. Mohr, M. Rayet, H. Utsunomiya, and A. Zilges}
\begin{frontmatter}
\title{Direct Determination of Photodisintegration Cross Sections and
  the p-process}
\author[1]{H. Utsunomiya}
\author[2]{P. Mohr}
\author[3]{A. Zilges}
\author[4]{M. Rayet}

\address[1]{ Department of Physics, Konan University,\\
 8-9-1 Okamoto, Higashinada, Kobe 658-8501, Japan }
\address[2]{ Strahlentherapie, Diakoniekrankenhaus Schw\"abisch Hall, \\ D--74523 Schw\"abisch Hall, Germany }
\address[3]{ Institut f\"ur Kernphysik, Technische Universit\"at Darmstadt,
 \\ Schlossgartenstra{\ss}e 9, D-64289 Darmstadt, Germany }
\address[4]{ Institut d'Astronomie et d'Astrophysique, Universit$\acute{e}$ Libre de Bruxelles,
\\ Campus de la Plaine, CP226, 1050 Brussels, Belgium.}
\begin{abstract}
Photon-induced reactions play a key role in the nucleosynthesis of
heavy neutron-deficient nuclei, the so-called p-nuclei. In this
paper we review the present status of experiments on
photon-induced reactions at energies of astrophysical importance
and their relevance to p-process modeling.
\end{abstract}
\end{frontmatter}

\section{Introduction}
\label{sec:intro}
The p-nuclei refer to stable, heavy nuclides that are
neutron-deficient and can not, for that reason, be produced in
stars by the slow or rapid neutron capture chains (s- or
r-processes), unlike the majority of heavy nuclei with charge
number in excess of the value $Z=26$ (Fe). Thirty-five nuclei are
classically considered as p-nuclei, with $Z$ ranging from 34 (Se)
to 80 (Hg), although 5 of them can also be produced to some extend
by the s-process. All p-nuclei can be synthesized from the
destruction of pre-existing seed nuclei of the s- and r-type by a
combination of \rpg\ captures and \rgn , \rgp\ or \rga\
photoreactions. Complemented by some $\beta^+$, electron captures
and \rng\ reactions, those nuclear flows are referred to as the
p-process. That p-nuclei are produced from existing s- or r-seed
nuclei is comforted by the fact that in the solar system they
represent only a small fraction (0.01 to 1\%, exceptionally of the
order of 10\%), of the isotopic content of the corresponding
elements.

The \rpg\ reactions require both high temperatures and large
proton densities and appear to  contribute only and probably
marginally, to the production of the lightest p-nuclei.
Photodisintegrations are thus expected to play the leading role in
the p-process. Temperatures in excess of about $T_9 = 1.5 $
($T_9=T/10^9$ K, where T is the temperature in Kelvin) are
required for photodisintegrations to take place on time scales
comparable to stellar evolutionary ones, and may not exceed $T_9 =
3.5 $ in order to avoid the photoerosion of all the heavy nuclei
to the more stable nuclei of the ``iron peak". It is also
necessary to freeze-out the photodesintegrations on a short enough
time-scale, typically of the order of one second. Those
constraints are nicely satisfied in the deep O-Ne-rich layers of
massive stars exploding as type II supernovae (SNe-II). The SN-II
is undoubtedly the most studied and the most
satisfactory scenario for the p-process \cite{woo78,ray95,rau02}.
Other plausible sites for the p-process, like pre-supernova
burning phases of massive stars or the explosion of type Ia
supernovae, have also been explored (see \cite{argo03} for a very
complete review of those works).

In order to estimate the number of photoreactions per unit of time
in a given volume of a star at temperature $T$, one has to
integrate over the energy $E$ the cross section $\sigma (E)$
weighted by the photon energy distribution $n_\gamma(E,T)$ times the speed of light $c$. As Sect.\ref{sec:basic} will show more quantitatively, only the high
energy tail of $n_\gamma(E,T)$ contributes to the rate, the
integrand being non-negligible only on a relatively narrow window
of photon energies. Knowing $n_\gamma(E,T)$, one might expect
photodisintegration rates to be determined easily by the
measurement of the cross section on that energy range, typically a
few MeV in the 1--10 MeV domain (Sect.~\ref{sec:basic}).

However, direct determinations of reaction rates for the p-process
suffer from two major limitations. The first is the fact that the
p-process involves thousands of photoreactions (not to speak of
the secondary nuclear transmutations mentioned above) and that
most of the involved nuclei are unstable, which means that only a
tiny fraction of those reactions can be measured in the
laboratory. The second limitation is that in a gas at high
temperature, excited levels of the target nuclei are populated
according to the Boltzmann statistics, so that photoreactions on
excited levels must be taken into account. This thermalization
effect is specially important here because of the high
temperatures involved in the p-process and because
photodisintegrations are specially sensitive to threshold effects.
This is illustrated in Sect.~\ref{sec:brems} in the case of \rgn\
reactions.

If the direct determination of astrophysical rates at work in the
p-process is clearly out of reach, experimental studies of
photodisintegration cross sections in the relevant energy range
and for nuclei as close as possible to the neutron deficient side
of the valley of stability are of crucial importance to test the
nuclear reaction models used to calculate the rates. Valuable
pieces of information are also obtained, more traditionally, by
the measurement of cross sections in the reverse, radiative
capture channel. They can indeed be used to constrain the
calculation of the rates in the photoreaction channel via the
reciprocity theorem. One must keep in mind however that such
measurements are but a fragment of the information needed for the
calculation of the reverse rate. To be correct, such a calculation
requires the knowledge of all non negligible cross sections from
any excited state of the target nucleus to any state of the
residual one.

Direct measurements of photodisintegration cross sections
constitute therefore an independent set of data and the most
straightforward way to constrain the calculation of the
corresponding astrophysical rate. Real-photon source facilities
have been developed and the interest of some of
these facilities for the study of the p-process is discussed in
the present paper\footnote{In this paper, we exclude
virtual-photon sources like electron scattering and Coulomb
excitation.}. After generalities on photon-induced reactions in
Sect.~\ref{sec:basic}, Sect.~\ref{sec:brems} presents some results
for \rgn\ reactions obtained at the bremsstrahlung facility of the
Technische Universit\"at Darmstadt. It is a nice feature of
bremsstrahlung facilities that they can be tuned to produce photon
spectra which approximate the high energy part of the Planck
spectrum for temperatures of interest for the p-process. The
obtained experimental rate can be directly compared to the
calculated rate for photoreaction on a ground-state nucleus.

Quasi-monochromatic photon beams with tunable energy can be
obtained using the technique of laser inverse-Compton scattering.
This technique has been successfully applied recently to nuclear
astrophysics at the National Institute of Advanced Industrial
Science and Technology in Japan and is described in
Sect.~\ref{sec:LCS}. The excitation function provided by such
experiments is extremely useful information to check
theoretical models of nuclear reactions. Section~\ref{sec:gamgam}
presents some results of ($\gamma,\gamma'$) experiments, providing direct
insight into the $\gamma$-strength below the particle threshold.
All the experiments described in this paper shed new light on the
low-energy tail of the $\gamma$-ray strength in nuclei and are
therefore expected to improve the rate predictions.

Improving the theoretical predictions for photodisintegration
rates will put p-process nucleosynthesis calculations on a firmer
ground. As mentioned before the p-process takes place at very high
temperatures but the nuclei involved are in a region of the
nuclear chart where basic quantities like masses or $\beta$-decay
rates are either known or rather reliably estimated. Experimental
data on photoreaction cross sections, even scarce, are therefore a
very precious ingredient to test the validity of Hauser-Feshbach
cross section calculations in the nuclear region of interest.
Sect.~\ref{sec:p-proc} briefly discusses the impact of such data
on the production of the p-nuclei, in relation with the
uncertainties inherent to the envisioned astrophysical scenarios.
Sect.~\ref{sec:p-proc} also discusses a few examples where the
production of a p-nuclide is directly related to the measurement
of specific photonuclear cross sections.

Finally, perspectives for new experimental techniques and
measurements related to astrophysical problems are 
presented in Sect.~\ref{sec:outlook} with emphasis on an
insertion light source of the SPring-8, a synchrotron radiation 
facility of the third generation.  Conclusions are drawn in Sect.~\ref{sec:concl}.

\section{Basic considerations on photon-induced reactions}
\label{sec:basic}

\subsection{Photoreactions on nuclei in the ground state}

The reaction rate $\lambda_{(\gamma,j)}$ for a photoreaction
induced on a ground-state nucleus, leading to the emission of
particle $j$ is given by the expression :
\begin{equation}
\lambda_{(\gamma,j)} =
  \int_0^\infty
  c \,\, n_\gamma(E) \,\, \sigma_{(\gamma,j)}(E) \,\, dE
\label{eq:rate}
\end{equation}
where $c$ is the speed of light, $\sigma_{(\gamma,j)}$ the cross
section and $n_\gamma(E)$ the number of photons per unit volume
and energy $E$. In a stellar interior at a temperature $T$,
$n_\gamma(E)$ is remarkably close to a black-body or Planck
distribution:
\begin{equation}
n_\gamma(E,T) \, dE =
    \frac{1}{\pi^2} \, \frac{1}{(\hbar c)^3} \,
    \frac{E^2}{\exp{(E/kT)} - 1} \, dE \quad \quad \quad .
\label{eq:planck}
\end{equation}
It has its maximum at energies around $E \approx\frac{3}{2} kT$
which is of the order of a few hundred keV in the temperature
range $1.5 \le T_9 \le 3.5$ ($T_9 = 1$ corresponds to $kT =
86$\,keV). At energies of several MeV the photon density is
governed by the exponential decrease in Eq.~\ref{eq:planck}.

If Eq.~\ref{eq:planck} is substituted for $n_\gamma(E)$ in
Eq.~\ref{eq:rate}, the rate becomes a function of the parameter
$T$. It is then possible to define the photon energy range which
is the most relevant for determining $\lambda_{(\gamma,j)}(T)$ at
the temperatures of astrophysical interest. This results from the
properties of the integrand of Eq.~\ref{eq:rate} which differs
significantly from zero only in a relatively small energy range.
We call this range the Gamow window by reference to the case of
reactions induced by charged particles in a thermalized
environment, where the entrance channel energy distribution is
given by the Maxwell-Boltzmann statistics. Here however one has to
distinguish between \rgn\ reactions, where the position of the
Gamow window is determined by the reaction threshold, and \rgp\ or
\rga\ reactions where it is shifted and broadened by the Coulomb
barrier in the outgoing channel \cite{pm719}.

\begin{figure}
\begin{center}
\includegraphics[ bb = 96 135 492 342, width = 140 mm, clip]{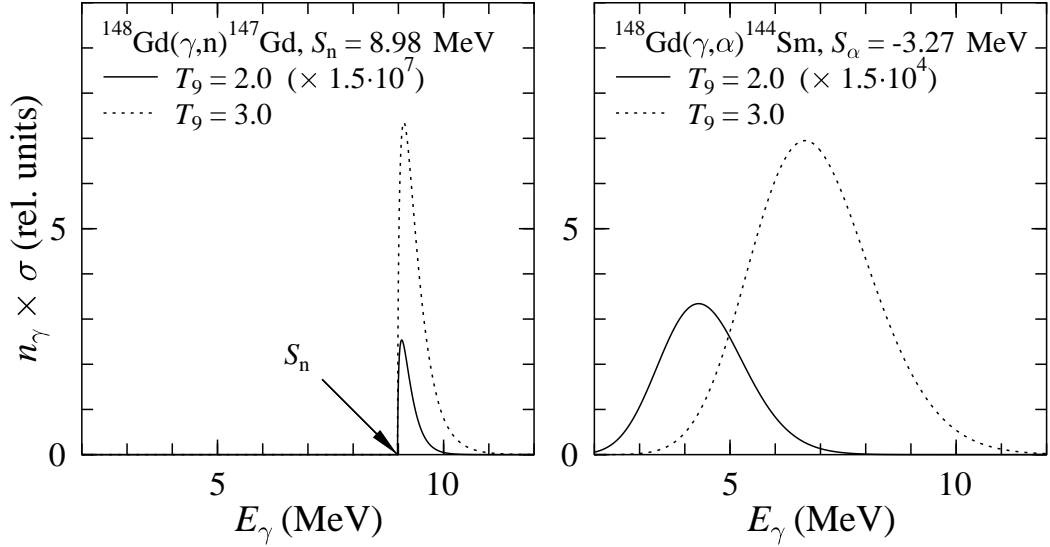}
\end{center}
\caption{
  Gamow window (integrand of Eq.~\ref{eq:rate}) for \rgn\
  (left panel) and \rga\ reactions (right panel) on the ground state of the target
  nucleus $^{148}$Gd, with separation energies
  $S_{\rm{n}} = 8.98$\,MeV and $S_\alpha = -3.27$\,MeV.
  Note the temperature dependence of the position of the Gamow window when
  the emitted particle is charged as well as the significant changes of the
  vertical scales in both panels when $T_9$ goes from 2 to 3.
  }
\label{fig:gamow}
\end{figure}

The \rgn\ cross section close above the threshold $S_n$ can be
expressed \cite{wig48} as
\begin{equation}
\sigma_{(\gamma,{\rm{n}})}(E) = \sigma_0 \times
  \left(\frac{E-S_{\rm{n}}}{S_{\rm{n}}}\right)^{\ell+1/2}
\label{eq:thres}
\end{equation}
where $\ell$ is the angular momentum of the emitted neutron.
Eqs.~\ref{eq:planck} and \ref{eq:thres} locate the maximum of the
integrand in Eq.~\ref{eq:rate} at $E = S_{\rm{n}} +
kT/2$ for $\ell = 0$ \cite{pm488,kv63,pm688}. The corresponding
narrow Gamow window is shown in Fig.~\ref{fig:gamow} (left panel)
for the temperatures $T_9 = 2.0$ and 3.0. The Gamow window for
\rgn\ reactions thus remains close to the reaction threshold for
all relevant temperatures. Because of the strong temperature
dependence of the photon density, the reaction rate depends
sensitively on the temperature; it increases by a factor $6.7
\cdot 10^7$ from $T_9 = 2.0$ to 3.0 for the example shown in
Fig.~\ref{fig:gamow}.

For the photon-induced emission of charged particles we make the
assumption that the astrophysical $S$-factor is roughly constant
for the inverse capture reaction. This leads to a Gamow window for
\rga\ and \rgp\ reactions which peaks at the sum of the separation
energy $S_{\alpha,{\rm{p}}}$ and the Gamow window peak energy
$E_0$ of the inverse capture reaction,
\begin{equation}
E_\gamma = S_{\alpha,{\rm{p}}} + E_0 \quad ,
\label{eq:gamow}
\end{equation}
and has the same width as the Gamow window of the capture
reaction. In contrast to the \rgn\ reaction, the position of the
Gamow window for \rga\ and \rgp\ reactions changes significantly
with temperature (see right panel of Fig.~\ref{fig:gamow}).
However, the temperature dependence of the reaction rate is less
strong than in the \rgn\ case. From $T_9=2$ to 3 it changes by a
factor of $4.4 \cdot 10^4$, i.e. three orders of magnitude less
than in the neutron case. Consequently, the branching ratio
between \rgn\ and \rga\ reactions will depend sensitively on the
temperature. This remark, which applies also to some extend to the
\rgp\ reactions, is of particular importance for the path followed
by the nuclear flow in the p-process (see e.g. \cite{rpa90}).

We remark also that in the nuclear mass region of interest for the
p-process and not too far from stability, \rga\ reactions usually
have much larger reaction rates than \rgp\ reactions because in
that region $\alpha$ particles have small or even negative binding
energies, whereas typical proton separation energies are of the
order of several MeV. Therefore, when the $\alpha$ particle is
replaced by a proton, the smaller Coulomb barrier, reducing $E_0$
in Eq.~\ref{eq:gamow} by a factor $2^{4/3}$, does not compensate
the strong increase in binding energy ($S_p-S_\alpha$) and the
\rga\ reaction operates at significantly lower photon energies,
and thus higher photon densities, than the \rgp\ reaction.

Very few experimental data on photodisintegration cross sections
are available in the literature at the energies of interest for
the p-process. For \rgn\ reactions most of the data have been
measured around the giant dipole resonance (GDR) with high
precision; however, close to the threshold, the data have
typically much larger uncertainties \cite{Die88}. The situation is
even worse for \rga\ reactions, for which the rare available
measurements have all been made at energies much higher than the
Gamow window (see e.g. \cite{Ant91}). Another difficulty comes
from the fact that the p-nuclei are very little abundant naturally 
and that experiments on those nuclei usually require targets made of a
considerable amount of highly enriched material.

\subsection{Thermalization effect under stellar conditions}

In stellar environments nuclear excited states are thermally
populated.  At the high temperatures we consider, thermalization
may enhance the photoreaction rates by several orders of
magnitudes (examples are found in Sect.~\ref{sec:brems}). The
right hand side of Eq.~\ref{eq:rate}, which corresponds to
photodisintegration from the ground state only, must then be
replaced by a sum of the rates $\lambda^{\mu}_{(\gamma,j)}(T)$ for
photodisintegration from all (ground and excited) states $\mu$,
each term being weighted by the appropriate Boltzmann factor. The
true astrophysical rate $\lambda^*$ is therefore defined  by
%
\begin{equation}
\lambda^*_{(\gamma,j)}(T) = \frac{1}{G(T)}\sum_\mu
\frac{(2J^\mu+1)}{(2J^0+1)}\lambda^{\mu}_{(\gamma,j)}(T)~
\exp{(-\varepsilon^\mu/kT)}\quad , \label{eq:astrorate}
\end{equation}
%
where, in $\lambda^{\mu}_{(\gamma,j)}(T)$,
$\sigma_{(\gamma,j)}(E)$ is replaced by the cross section
$\sigma^{\mu}_{(\gamma,j)}(E)$ for photodisintegration from state
$\mu$ and where $G(T)=\sum_{\mu}{(2J^{\mu}+1)}/{(2J^{0}+1)}
\exp(-\varepsilon^{\mu}/kT)$ is the temperature-dependent
normalized partition function of the target nucleus.

Although the gross effect of thermalization is a shift of the
Gamow window to lower photon energies by some mean excitation energy, it
is mandatory to perform the sum explicitly (or, if necessary, to
integrate on a model level density for the target), when more than
one excited level is populated, which is most often the case for
relatively massive nuclei away from shell closures 
in the considered temperature range.

Clearly the multitude of astrophysical rates needed to describe
the nucleosynthesis of the p-nuclei has to be calculated
theoretically. In all existing p-process calculations, reaction
rates are calculated in the framework of the Hauser-Feshbach (HF)
statistical model. We refer to \cite{HWFZ76,argo03,Rau00} for a
description of the HF model and of its underlying hypothesis. Let
us just recall here that the HF cross section for the reaction
$I^{\mu} + j\rightarrow L^{\nu} + k$ where particle $j$ is
captured on nucleus $I$ in excited state $\mu$, leaving residual
nucleus $L$ in state $\nu$ and particle or photon $k$, is obtained
from the transmission coefficients for the formation of or decay
from all states $J^\pi$ of the compound nucleus which can be
formed from the quantum numbers of the entrance channel. Since in
stellar conditions the target nucleus is in thermal equilibrium it
can be shown \cite{HWFZ76} that the astrophysical rates for the
forward and reverse channels of the reaction $I + j\rightarrow L +
k$ are symmetrical and therefore obey reciprocity, which is not
the case when the target nucleus is in its ground state only. This
remains true when $k$ is a photon, so that the photodisintegration
rate of a nucleus $L$ leading to particle $j$ and residual nucleus
$I$ is directly proportional to the radiative capture rate of
particle $j$ on $I$, $N_A\langle\sigma v\rangle^*_{(j,\gamma)}$,
where * means as before that the rate takes target thermalization
into account.

We want to emphasize here that as long as the conditions for the
application of a statistical model for the reaction cross sections
are met, which is the case to a large extend for typical p-process
nuclear flows as discussed in Sect.~\ref{sec:p-proc}, the
uncertainties involved in any HF cross section calculation are
essentially related to the evaluation of the nuclear quantities
necessary for the calculation of the partition functions $G(T)$ as
well as the transmission coefficients entering the calculation of
$\langle\sigma v\rangle^*_{(j,\gamma)}$. Not only the ground state
properties (masses, deformations, matter densities) of the target
and residual nuclei have to be known, or, when not available
experimentally, have to be obtained from nuclear mass models, but
the excited state properties are also indispensable. Experimental
data may be scarce, especially for nuclei located far from the
valley of nuclear stability and frequent resort to a level density
prescription is mandatory. The transmission
coefficients for particle emission are calculated by solving the
Schr\"odinger equation with the appropriate optical potential for
the particle-nucleus interaction. The case of the $\alpha$-nucleus
potential is of particular significance for the p-process but
suffers from the scarcity of $\alpha$-nucleus cross section
measurements at sub-Coulomb energies, especially for $A>100$
nuclei, and from the difficulties to construct theoretically
global and reliable $\alpha$-nucleus potentials (see recent
attempts in \cite{pm2000,demet2002} and \cite{argo03} for a review
of the present situation).

In order to obtain reliable predictions for the astrophysical
rates, a strict methodology is compulsory: calculated nuclear
properties, constrained by indispensable but sporadic experiments,
should rely on coherent sets of data, based whenever possible on
microscopic models of the nucleus. Extensive work along this line
has been performed in the framework of the on-line library BRUSLIB
\cite{sgtours03} \footnote{accessible at url
http://www.astro.ulb.ac.be}.

The photon transmission function requires a particular attention
in the case of photonuclear reactions and is calculated assuming
the dominance of dipole E1 transitions. Reaction theory relates
the $\gamma$-transmission coefficient for excited states to the
ground state assuming the GDR is built on each excited state and
has a Lorentzian representation, at least for medium- and
heavy-mass nuclei. Experimental photoabsorption data confirm the
simple semi-classical prediction of a Lorentzian shape at energies
around the resonance energy, but this description is less
satisfactory at lower energies, and especially near the reaction
threshold. Even if a direct knowledge of the astrophysical rate of
Eq.~\ref{eq:astrorate} is not accessible to experiments,
photoreaction cross section measurements in the Gamow window
energy range will be extremely useful to improve our knowledge of
the dipole strength functions at low energy.
%
%

\section{Photodisintegration measurements with bremsstrahlung}
\label{sec:brems}
Bremsstrahlung facilities provide intense photon radiation with
energies up to the energy of the incoming electron beam. As shown
in Sect.~\ref{sec:basic}, the astrophysically relevant energy
range is located at several MeV. Hence facilities with electron
energies around 10\,MeV are best suited for experiments of
astrophysical interest.


\begin{figure}
\begin{center}
\includegraphics[ bb = 33 113 574 694, width = 140 mm, clip]{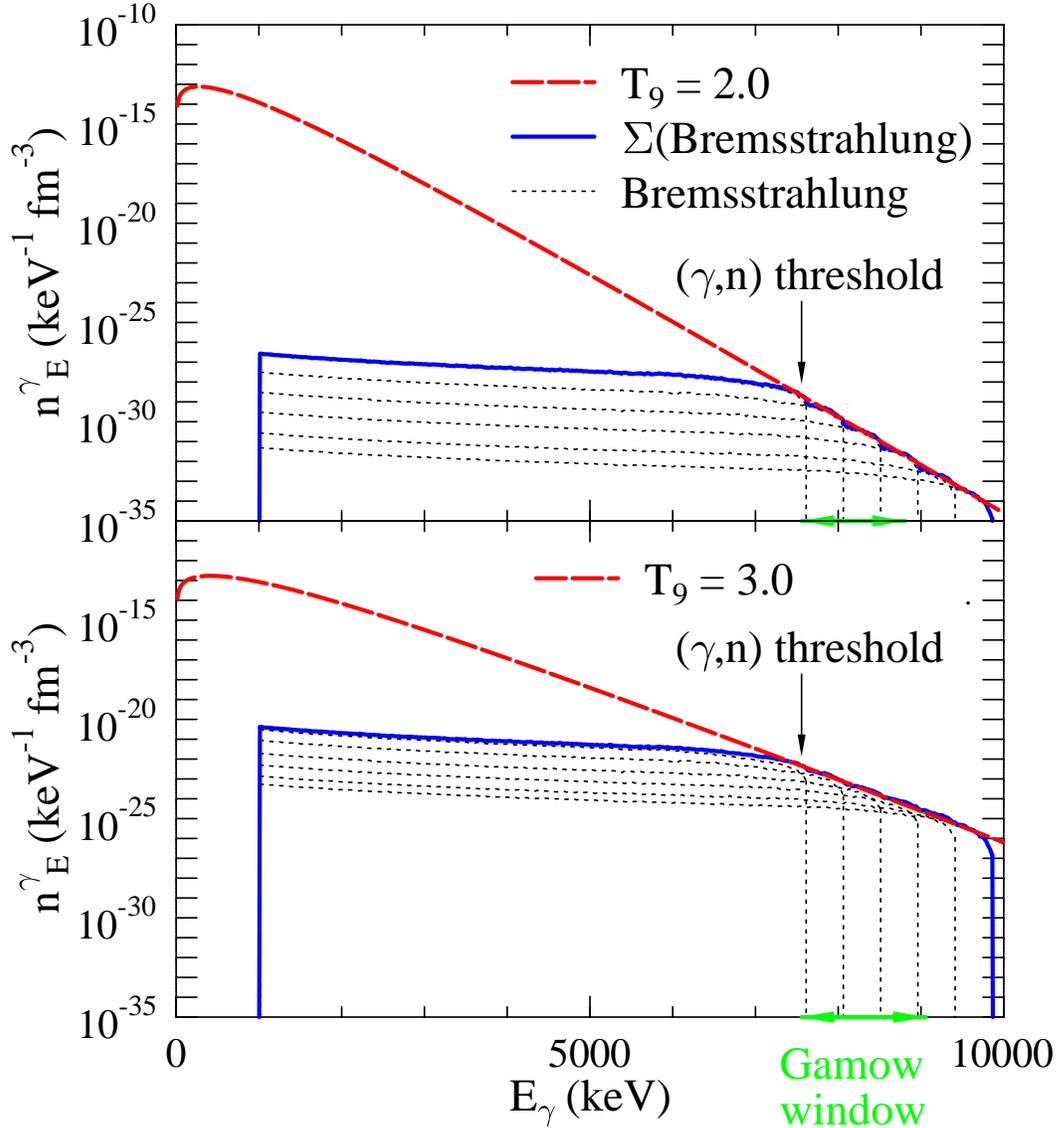}
\end{center}
\caption{ Approximation of the thermal photon energy distribution
  (dashed line) by a weighted sum of the end-point portions of bremsstrahlung
  spectra (full line) for $T_9 = 2$ (upper panel) and $T_9 = 3$ (lower panel),
  in the corresponding Gamow window, located close above the \rgn\ threshold.
  The individual bremsstrahlung spectra are shown with dotted lines.
}
\label{fig:approx}
\end{figure}

The photoactivation technique, obviously limited to reactions
where an unstable nucleus is produced, has been chosen for the
experiments using bremsstrahlung because a direct detection of the
emitted particle is difficult in the huge bremsstrahlung
background.  Additionally, the excellent energy resolution of
$\gamma$-ray detectors allows a clear detection of the individual
reaction even in a chemically mixed target with natural isotopic
composition. For example, a clear photoactivation signal has been
observed for the $^{190}$Pt\rgn $^{189}$Pt reaction with a
$^{190}$Pt mass of about 100\,$\mu$g (i.e. 0.014\,\% natural
abundance of $^{190}$Pt in a 800\,mg natural platinum target). The
disadvantage of a low overall detection efficiency, which is of
the order of a few per cent in the best cases, is compensated by
the huge number of photons in the bremsstrahlung beam. This
results in typical measuring times of a few days.
\begin{table}
\caption{ Experimental results  for \rgn\ reaction rates,
$\lambda^{\rm{g.s.}}_{\rm{exp}}$, for $T_9 = 2.5$ (experimental
uncertainties are given in parentheses). They are compared to
theoretical predictions for photodisintegration from the target
ground state: $\lambda^{\rm{g.s.}}_{\rm{th}}$(M) and
$\lambda^{\rm{g.s.}}_{\rm{th}}$(MST) are calculated with the code
MOST (see text) and $\lambda^{\rm{g.s.}}_{\rm{th}}$(NS) is the
prediction of the code Non-Smoker. Additionally the rate
$\lambda^\ast$ for a thermalized target and the corresponding
enhancement factor
$\lambda^{\ast}/\lambda^{\rm{g.s.}}_{\rm{th}}$(NS) are also shown.
All rates are in s$^{-1}$.} \label{tab:results}
\begin{center}
\begin{tabular}{c c c c c c c}
\hline nucleus
&$\lambda^{\rm{g.s.}}_{\rm{exp}}$
&$\lambda^{\rm{g.s.}}_{\rm{th}}$(M)
&$\lambda^{\rm{g.s.}}_{\rm{th}}$(MST)
&$\lambda^{\rm{g.s.}}_{\rm{th}}$(NS)
&$\lambda^{\ast}$
&$\lambda^{\ast}/\lambda^{\rm{g.s.}}_{\rm{th}}$(NS) \\
\hline
$^{186}$W
& $3.1(4) \cdot 10^{2}$ & 1.1--2.8$\cdot 10^2$ & $2.5\cdot 10^2$ &
$2.6 \cdot 10^{2}$
& $1.0 \cdot 10^{5}$  & $4.0 \cdot 10^{2}$ \\
$^{185}$Re
& $1.9(7) \cdot 10^{1}$ & 1.0--4.7$\cdot 10^1$ & $4.4 \cdot 10^1$ &
$1.9 \cdot 10^{1}$ & $2.5 \cdot 10^{4}$
& $1.3 \cdot 10^3$ \\
$^{187}$Re
& $7.6(7) \cdot 10^{1}$ & 1.9--8.2$\cdot 10^1$ & $7.0\cdot 10^1$
&$7.2 \cdot 10^{1}$ &$8.4 \cdot 10^{4}$ & $1.2 \cdot 10^3$ \\
$^{190}$Pt
& $4(2) \cdot 10^{-1}$ & 1.1--4.8 $\cdot 10^{-1}$ & 2.9
$\cdot 10^{-1}$ & $1.8 \cdot 10^{-1}$ & $1.0 \cdot 10^{3}$
& $5.5 \cdot 10^{3}$ \\
$^{192}$Pt
&$5(2) \cdot 10^{-1}$ & 0.2--1.3$\cdot 10^0$ & 5.6 $\cdot 10^{-1}$
& $5.8\cdot 10^{-1}$& $1.9 \cdot 10^{3}$  & $3.3 \cdot 10^{3}$ \\
$^{198}$Pt
& $8.7(2) \cdot 10^{1}$  & 0.34--1.3 $\cdot 10^2$ & 1.1
$\cdot 10^2$ & $5.0 \cdot 10^{1}$
& $1.5 \cdot 10^{4}$  & $3.1 \cdot 10^{2}$ \\
$^{197}$Au
& $6.2(8) \cdot 10^{0}$  & 2.7--9.1 $\cdot 10^0$& 5.6 $\cdot 10^0$ &
$4.8 \cdot 10^{0}$& $5.1 \cdot 10^{3}$  & $1.1 \cdot 10^{3}$ \\
$^{196}$Hg
&$4.2(7) \cdot 10^{-1}$& 2.0--7.5 $\cdot 10^{-1}
$& 5.8 $\cdot
10^{-1}$ & $3.2 \cdot 10^{-1}$
& $5.4 \cdot 10^{2}$  & $1.7 \cdot 10^{3}$ \\
$^{198}$Hg
& $2.0(3) \cdot 10^{0}$  & 0.77--3.0 $\cdot 10^0$ & 2.1 $\cdot 10^0$
& $1.4 \cdot 10^{0}$
& $1.0 \cdot 10^{3}$  & $7.5 \cdot 10^{2}$ \\
$^{204}$Hg
& $5.7(9) \cdot 10^{1}$ & 0.47--1.9 $\cdot 10^2$ & 1.7 $\cdot 10^2$ &
$7.3 \cdot 10^{1}$& $3.1 \cdot 10^{3}$  & $4.3 \cdot 10^{1}$ \\
$^{204}$Pb
& $1.9(3) \cdot 10^{0}$ & 0.98--3.8 $\cdot 10^0$ & 3.0 $\cdot 10^0$ &
$1.5 \cdot 10^{0}$
& $2.5 \cdot 10^{2}$  & $1.6 \cdot 10^{2}$ \\
\hline
\end{tabular}
\end{center}
\end{table}

The general set-up of photoactivation experiments is simple. Here
we describe the set-up used at the TU
Darmstadt \cite{pm423}. Electrons with energies up to 10\,MeV and
currents up to about 50\,$\mu$A are provided by the
superconducting linear accelerator S-DALINAC. The electron beam is
completely stopped in a massive copper radiator. The
bremsstrahlung is collimated and hits the target roughly 1.5\,m
behind the radiator. For normalization, the incoming photons are
monitored by photon scattering in the $^{11}$B\rgg $^{11}$B$^\ast$
reaction (see Sect.~\ref{sec:gamgam}). Typical photon intensities
are of the order of $10^4 - 10^5$ keV$^{-1}$\,cm$^{-2}$\,s$^{-1}$. Alternatively
a relative measurement can be carried out using a standard with
well-known cross section and suitable properties for
photoactivation (low photoneutron threshold, reasonable half-life,
strong $\gamma$-ray lines after $\beta$-decay, high natural
abundance). The $^{197}$Au\rgn $^{196}$Au reaction has
already been used successfully \cite{kv707}, and measurements have
been performed to establish $^{187}$Re as another standard
\cite{mue04}. Excellent agreement has been found between
experiments using bremsstrahlung/photoactivation and monochromatic
photons/direct neutron detection, for both standard nuclei
$^{197}$Au and $^{187}$Re. A relative measurement can also be made
putting the target very close to the radiator, where the photon
intensity is roughly a factor of 300 higher than at the position
behind the collimator. After photoactivation the $\gamma$-rays
from the decay of the produced nuclei are measured using a
high-purity germanium detector.

The analysis of bremsstrahlung data is complicated because of the
broad energy spectrum of the bremsstrahlung photons. The experimental
yield $Y$ is proportional to
\begin{equation}
Y \sim\ \int_{S_{\rm{n}}}^{E_0} \sigma(E) \, N_\gamma(E,E_0) \,
dE\quad ,
\label{eq:yield}
\end{equation}
where $N_\gamma(E,E_0)$ is the number of photons per keV and
cm$^2$ in the brems\-strahlung spectrum with endpoint energy
$E_0$. Unfolding procedures have been used to extract the cross
section $\sigma(E)$, but such procedures are limited by
significant systematic errors. Alternatively, a reasonable
assumption on the threshold behavior or a theoretical prediction
for $\sigma(E)$ can be used to solve the integral in
Eq.~\ref{eq:yield}; in this case the experimental result is just
a normalization factor for the theoretical prediction.

Recently, a new method has been established to derive the ground
state reaction rates $\lambda$ by approximating the black-body
photon density $n_\gamma(E,T)$ (Eq.~\ref{eq:planck}) by a weighted
sum of bremsstrahlung spectra with different endpoint energies
$E_{0,i}$\ :
\begin{equation}
  c \, n_\gamma(E,T) \approx \sum_i a_i(T) \, 
  N_\gamma (E,E_{{0,i}})\quad ,
\label{eq:approx}
\end{equation}
where $a_i(T)$ is a set of weighting coefficients for a given
value of $T$. The excellent agreement between the thermal
distribution and the weighted sum in the relevant energy window
close above the threshold is shown in Fig.~\ref{fig:approx} for
$T_9 = 2.0$ and $T_9 = 3.0$; hence the weighted sum of
bremsstrahlung spectra may be called a ``quasi--thermal'' spectrum
with variable temperature.
Up to now results have been obtained for a number of nuclei which
are listed in Tab.~\ref{tab:results} \footnote{Minor differences
between the values in Tab.~\ref{tab:results} and already published
numbers come from an improved analysis of the shape of the
bremsstrahlung spectra close to the endpoint energy.}.
The values $\lambda^{\rm{g.s.}}_{\rm{exp}}$ are derived from
bremsstrahlung experiments with a superposition of spectra
(Eq.~\ref{eq:approx}) corresponding to a temperature $T_9=2.5$.
Those values are compared to theoretical predictions for
photodisintegration from the target ground state. Column
$\lambda^{\rm{g.s.}}_{\rm{th}}$(M) shows the minimum and maximum
values of the rates calculated with the code MOST \cite{go98b},
for 14 different sets of the nuclear data necessary to calculate
the HF cross sections, and $\lambda^{\rm{g.s.}}_{\rm{th}}$(MST)
are MOST ``standard" values (see \cite{argo03} and
\cite{sg_priv04} for Re).
The rates $\lambda^{\rm{g.s.}}_{\rm{th}}$(NS) were derived \cite{Raupc04} from cross sections calculated with the code Non-Smoker (such cross sections can be found in \cite{Rau04}). The importance of the target
thermalization is well illustrated in the last two columns which
show the rates $\lambda^\ast$ for a thermalized target as well as the corresponding enhancement factors \cite{Raupc04}.
It is clear that the thermalization effect depends strongly on the
detailed level structure of the target nucleus and cannot be
estimated by purely qualitative arguments.

Table~\ref{tab:results} shows that there is never a strong
disagreement between experimental data and theoretical predictions
and that no systematic trend for over- or under-estimate of the
experimental data can be traced from the present data on very
heavy nuclei. The experimental data lie
within the ranges of values spanned by the MOST rates obtained
with different but reasonable choices of nuclear physics data. The
ratio of the maximum to the minimum values never exceeds 6.5,
which is a rather favorable situation \cite{argo03}. However, the
extreme values do not necessarily correspond to the same set of
nuclear physics data so that the comparison made in
Table~\ref{tab:results} can not be used to discriminate between
the different nuclear physics ingredients used in the HF
calculations.

\section{Photodisintegration measurements with laser inverse-Compton scattering $\gamma$ rays}
\label{sec:LCS}

The inverse Compton scattering was first studied theoretically in
collisions of cosmic rays on thermal photons in space
\cite{feen48}.  The idea of producing $\gamma$ rays in the
laboratory by interactions between laser photons and relativistic
electrons was born in 1963 \cite{milb63,arut63}. Technical facets
of the idea for practical use were developed in the 1980's
\cite{fede80} but the application of this technique to
astronuclear physics had been ignored until recently.

Head-on collisions of laser photons with relativistic electrons
produce $\gamma$ rays with an energy given to an excellent
approximation by,

\begin{equation}
E_\gamma = \frac{4 \gamma^2 \varepsilon_L}{1 + (\gamma \theta)^2 +
4 \gamma \varepsilon_L/(m c^2)}, \label{eq:invcomp}
\end{equation}

\noindent where $\gamma = E_e/mc^2$, $E_e$ is the electron beam
energy and $m$ its rest mass, $\varepsilon_L$ is the laser photon
energy, and $\theta$ the scattering angle of laser photons with
respect to the electron beam.  Either a conventional laser or a
free-electron laser \cite{HIgS} can be employed. The angle $\theta
=0$ corresponds to the maximum of $E_\gamma$, as well as of the
cross section for photon scattering, according to the
Klein-Nishina formula. Also at that angle the polarization of the
laser photon is conserved. Collimating scattered photons at
$\theta \sim 0$ produces $\gamma$ rays with energy spread, $\Delta
E_\gamma/E_\gamma = [(2 \Delta E_e/E_e)^2 + (\gamma \Delta
\theta)^4]^{1/2}$\cite{sand83}. In practice, this energy spread is
determined by $\Delta \theta = (\theta_e^2 + \theta_c^2)^{1/2}$,
where $\theta_e$ is the electron beam divergence and $\theta_c$
the collimator half angle, rather than by the energy spread of the
electron beam $\Delta E_e /E_e$.  By changing either the electron
beam energy or the laser wavelength, the laser inverse-Compton
scattering plays the role of a {\it photon accelerator}, producing
a $\gamma$-ray beam that is energy-variable, quasi--monochromatic
and linearly- (or circularly-)100\% polarized.  That technique is
superior to the positron annihilation in flight because the latter
is beset with the positron bremsstrahlung
\cite{Die88}.

\subsection{Measurements with the LCS $\gamma$ beam at AIST}

Fine pencil-like beams (typically 2 mm in diameter) of $\gamma$
rays are available based upon the laser inverse-Compton scattering
(LCS) at the National Institute of Advanced Industrial Science and
Technology (AIST) \cite{ohga91}. Their production utilizes the
conventional lasers (Nd:YLF and Nd:YVO) in both Q-switch and CW
modes and electron beams in the storage ring TERAS.  The $\gamma$
energy is varied in the region of 1 - 40 MeV by tuning the
electron beam energy from 200 to 800 MeV.  An energy resolution of
1 - 10 \% in FWHM and nearly 100\% polarization are achieved.
Because of the monochromaticity, the LCS $\gamma$ beam is best
suited to excitation function measurements of photoneutron
reactions near threshold with enriched-target material of the
order of 1g. In addition, photo-activation of natural foils can be
done for nuclei whose isotopic abundance is sufficiently large.

The AIST-LCS $\gamma$-beam with a rather limited intensity
(10$^{4-5}$ photons/sec) has been used to measure cross sections
of $^{9}$Be($\gamma$,n)$\alpha \alpha$ of interest for the
nucleosynthesis in supernovae \cite{uts01,sum02}, of
$^{181}$Ta($\gamma$,n)$^{180}$Ta for the p-process
nucleosynthesis \cite{uts03}, and D($\gamma$,n)p for big bang
nucleosynthesis \cite{har03}.  More recently, photoneutron cross sections have
been measured on the $^{186}$W, $^{187}$Re, and $^{188}$Os nuclei,
of interest for s-process nucleosynthesis and cosmochronometry, as
well as on $^{93}$Nb and $^{139}$La for p-process studies. In
these studies, a 4$\pi$-type detection of neutrons was carried
out; the latest version of the neutron detector consists of double
rings with a total 16 $^{3}$He proportional counters embedded in a
polyethylene moderator with an overall detection efficiency up to
46\% depending on neutron energy.  The average energy of neutrons
emitted in photodisintegration of medium/heavy nuclei at a given
E$_{\gamma}$ was determined by the so-called ring ratio, the ratio
of neutrons detected by the inner and outer rings of 8 $^{3}$He
counters each.  Typical time for measuring the excitation function
over the Gamow window is 1 hour per energy, thanks to large GDR cross sections 
even in the tail region except at energies very close to neutron thresholds.
It is noted that the nuclear database of the electric giant dipole
resonance \cite{Die88} lacks sufficient accuracy in the energy
region of astrophysical importance as is evidenced by the
non-vanishing values of the cross sections below threshold.

\begin{figure}[htb]
\begin{center}
\includegraphics [width=18pc]{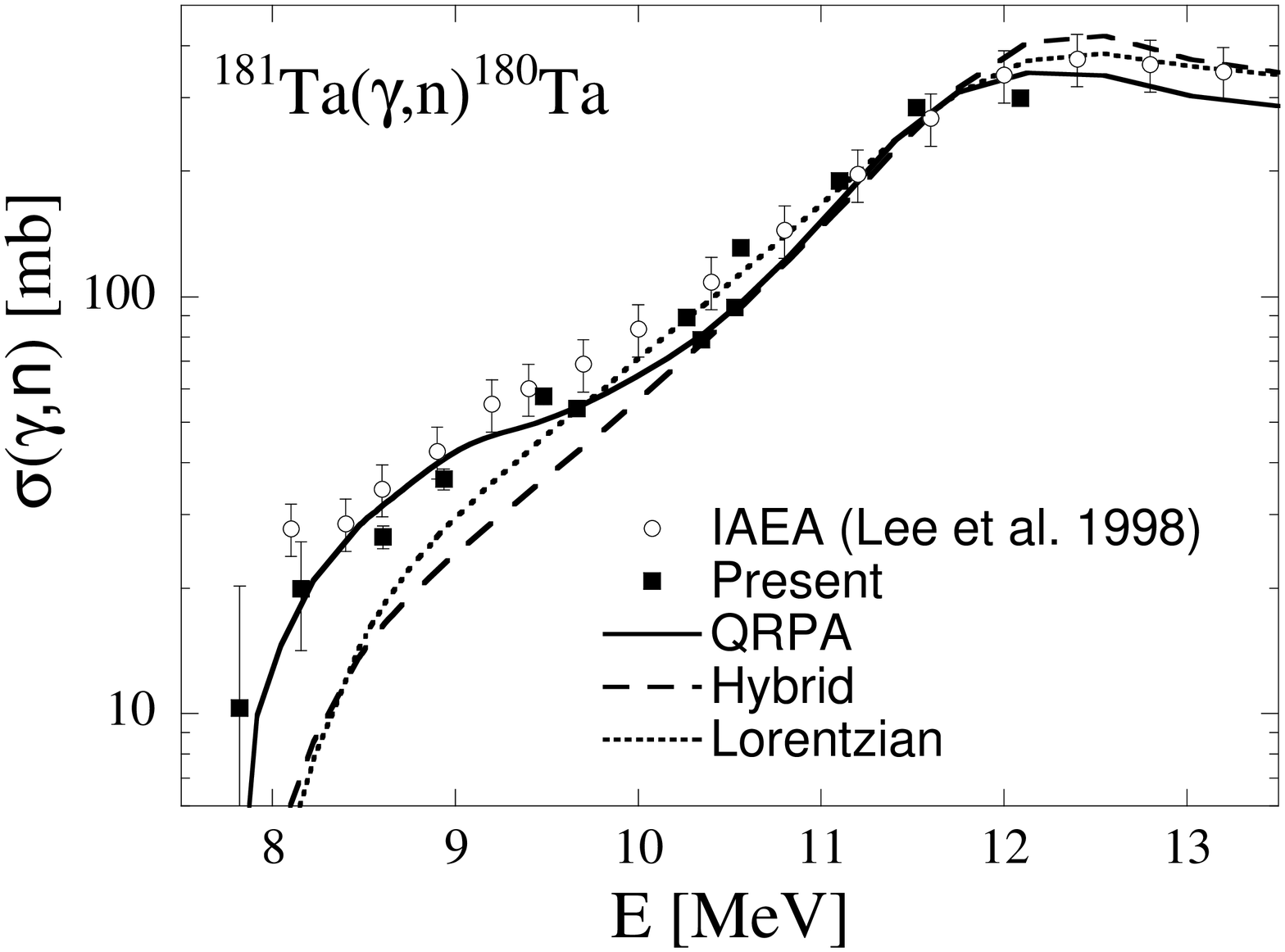}
\end{center}
\caption{ Photodisintegration cross sections for
$^{181}$Ta($\gamma$,n)$^{180}$Ta \cite{uts03}.}
\label{fig:tantalum}
\end{figure}

Figure~\ref{fig:tantalum} shows the experimental \rgn\ cross
section on $^{181}$Ta obtained with the LCS $\gamma$ beam
\cite{uts03}, compared with data recommended by IAEA on the basis
of former measurements. Those data provided constraints on the
low-energy tail of the dipole strength function.  Their
interpretation has necessitated a microscopic understanding of
threshold behavior of photoneutron cross sections, showing the
advantage of a QRPA calculation over a conventional Lorentzian- or
hybrid-model analysis. The stellar photoneutron rate for
$^{181}$Ta was calculated in the Hauser-Feshbach statistical model
with the QRPA result for the E1 strength. Those results have been
used in \cite{uts03} to re-examine the problem of $^{180}$Ta in
the p-process (see Sect.~\ref{sec:p-proc}). Nuclear challenges
remain in order to reliably evaluate the $^{180}$Ta p-process
yield. They include measurements of the $^{180}$Ta
photodestruction rate and the $^{181}$Ta photo-neutron branching
to the $^{180}$Ta ground and first excited states \cite{USK}. Such
information would also help constraining reaction models.

\subsection{Data reduction}
In nuclear astrophysics experiments, it is of crucial importance
to provide absolute cross sections with high accuracy.  Here we
describe a methodology for deducing reliable cross sections from
quasi--monochromatic photon-induced reactions.

The number of neutrons $n_{\rm{exp}}$ emitted in the
photodisintegration experiment is related to the ($\gamma$,n)
cross section $\sigma(E_\gamma)$ by its integral over the photon
energy distribution $n_\gamma(E_\gamma)$:
\begin{equation}
n_{\rm{exp}} = N_{\rm{T}}\, 
h\, 
\int{n_\gamma(E_\gamma)\, 
\sigma(E_\gamma) \, dE_\gamma},
\label{eq:sigma}
\end{equation}
where $N_{\rm{T}}$ is the number of target atoms per unit area,
and $h$ is the correction factor for a thick-target measurement, $
h = (1 - e^{- \mu t})/ \mu t$ (with target thickness $t$ and
attenuation coefficient of target material $\mu$).  It is obvious
that for ideally monochromatic photons with energy $E_0$, the
integral is replaced by $N_\gamma$ $\times$ $\sigma(E_0)$ where
$N_\gamma$ is the number of incident photons.

Let us write $\sigma(E_\gamma)$ as a Taylor series,
\begin{equation}
\sigma(E_\gamma) = \sigma(E_{0}) + \sigma^{(1)}(E_{0})(E_\gamma - E_{0}) \,
+ \frac{1}{2} \sigma^{(2)}(E_{0})(E_\gamma - E_{0})^{2} \,
+ \cdot \cdot \cdot,
\label{eq:taylor}
\end{equation}

\noindent where $\sigma^{(i)} = d^{i} \sigma(E)/dE^{i}$.  When the
average energy is chosen for $E_{0}$, putting the Taylor series
into the integral in Eq.~\ref{eq:sigma} yields
\begin{equation}
\int{n_\gamma(E_\gamma) \, 
\sigma(E_\gamma) \, dE_\gamma} = \,
N_\gamma \{ \sigma(E_{0}) + s_{2}(E_{0}) + s_{3}(E_{0}) + \cdot
\cdot \cdot \} ,
\label{eq:series}
\end{equation}

\noindent where $s_{2}(E_{0}) = \frac{1}{2} \sigma^{(2)}(E_{0}) \,
[ \overline{E_\gamma^{2}} - E_{0}^{2} ]$, $s_{3}(E_{0}) =
\frac{1}{6} \sigma^{(3)}(E_{0}) [ \overline{E_\gamma^{3}} - 3
E_{0} \overline{E_\gamma^{2}} + 2 E_{0}^{3} ]$, etc., with
$\overline{E_\gamma^{i}} = \int{n_\gamma(E_\gamma) E_\gamma^{i}} d
E_\gamma / N_\gamma$. Note that $E_0 = \overline{E_\gamma}$, so
that $s_1(E_0)$ vanishes.

Experimentally, from Eqs.~\ref{eq:sigma} and \ref{eq:series}, the
bracketed Taylor series in Eq.~\ref{eq:series} is deduced from the
numbers of neutrons ($n_{exp}$) and incident photons
($N_\gamma$), as well as from the target properties. In order to
obtain the cross section at the average $\gamma$ energy,
$\sigma(E_{0})$, one has then to subtract the higher order terms
$s_{2}$, $s_{3}$, etc.

Recently this procedure was exactly followed in the data reduction
for the photodisintegration of $^{186}$W \cite{186w}.  It was
found that the subtraction resulted in only a few \% increase in
$\sigma(E_{0})$ in the energy region of astrophysical relevance,
where the energy dependence of the cross sections is dominated by
the s-wave neutron emission ($\ell = 0$ in Eq.~\ref{eq:thres}).
Thus, the LCS $\gamma$ ray is very close to monochromaticity at
the energies of astrophysical interest.


\section{The photoresponse of atomic nuclei
below the neutron threshold}
\label{sec:gamgam}

Because the thermal population of nuclear levels in stellar
interiors has such a strong influence on the astrophysical
photodisintegration rates, experimental studies on the
photoresponse of nuclei below the particle threshold are of
crucial importance for developing reliable models for such rates.
In particular, a detailed knowledge about the structure of dipole
excitations below the particle threshold is an important input for
these models.

An ideal tool to investigate the photoresponse of nuclei below the
particle threshold is real photon scattering or nuclear resonance
fluorescence (NRF) \cite{Kne96}. A ``white'' bremsstrahlung photon
spectrum is produced by stopping an electron beam in a radiator
target. This photon beam hits the target material and induces
dipole and, to a lesser extent, quadrupole transitions to higher
lying states. The $\gamma$\ decay of these states back to the
ground and excited states is observed with high resolution
($\Delta E/E \simeq 0.1 \%$) HPGe semiconductor detectors.

The pure electromagnetic excitation mechanism allows one to derive
absolute transition strength or decay width of the excited states
without any model dependency. Due to the high sensitivity of
present set-ups one gets a rather complete picture of the dipole
and quadrupole strength distribution in stable nuclei. Very
recently the bremsstrahlung experiments have been complemented by
($\vec{\gamma},\gamma'$) experiments using a polarized
quasi--monochromatic photon beam from laser Compton backscattering
\cite{Pie02}. This method gives easy access to additional
observables like parities and weak decay branchings.

\begin{figure}
\begin{center}
\includegraphics [width=30pc]{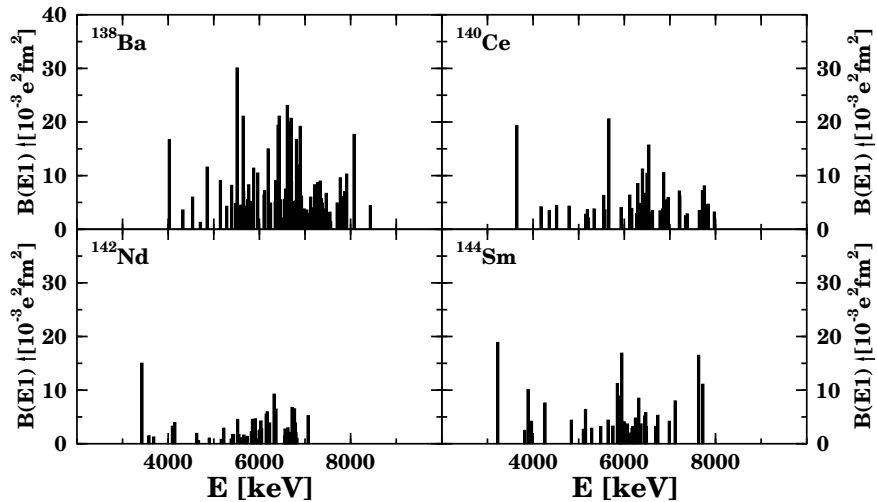}
\end{center}
\caption{
Distribution of electric dipole strength
in the $N=82$ isotones
$^{138}$Ba, $^{140}$Ce, $^{142}$Nd, and $^{144}$Sm.
}
\label{fig:e1distribution}
\end{figure}

The photoresponse of atomic nuclei is dominated by the giant
dipole resonance (GDR). However, recent measurements have shown
that a considerable part of the electric dipole strength remains
in the 1$\hbar\omega$\ region i.e.~around about 7 MeV in stable
nuclei \cite{Gov98,Her99,Zil02,Rye02} and is not shifted to the
GDR. In heavier nuclei this strength seems to be concentrated in a
resonance like distribution of 1$^{-}$\ states. This can be seen
in Fig.~\ref{fig:e1distribution} for the $N=82$ isotones. The
concentration of states between 6 and 8 MeV and the lack of
strength at higher energies shows that these are not just
statistical E1 excitations riding on the tail of the GDR but that
the states have their own nuclear stucture. The summed E1 strength
in this energy region exhausts up to one percent of the isovector
energy weighted sum rule. A similar concentration of E1 strength
has been found in lighter nuclei as well
\cite{Kne76,Bau00,Har02,Bab02}. 

In exotic nuclei with large neutron excess, up to 10\% of the
total isovector E1 strength has been found at very low energies,
both in photodissociation and in Coulomb excitation experiments in
inverse kinematics \cite{Lei01,Try03}. These experiments are a
useful complement to the high precision ($\gamma,\gamma'$)
experiments on nuclei in the valley of stability.

The structure of the E1 excitations is still
under intense discussion. Triggered by the
experimental observation that nuclei with
neutron separation energies of less than
S$_{n}$=10 MeV possess a neutron skin,
several models describe the mode as an
oscillation of this skin versus the proton/neutron
core \cite{Suz90,Cha94}. In addition several microscopic
calculations explain a considerable part of the E1 strength
as a dominant isoscalar mode \cite{Rye02,Vre02,Col00}.

Upcoming experiments will extend the systematics of the E1
strength distribution, will measure additional observables like
the isospin character and detailed branching ratios and will look
into the E1 response of exotic heavy nuclei. This will hopefully
allow a deeper insight into the structure of the E1 modes around
the  particle threshold and finally allow more reliable model
calculations of reaction rates with predictive power.

\section{Astrophysical p-process}
\label{sec:p-proc}
One question raised by the p-process studies is to know to what
extend the calculated p-nuclei abundances do reproduce the solar
system ones. A variety of explosive stellar sites in which matter
is heated to temperatures in the range $T_9 = 1.5$--3.5 succeed in
synthesizing p-nuclides with relative abundances in rough
agreement with the solar system isotopic content. The only serious
discrepancy concerns the large isotopic ratios of the Mo and Ru
p-nuclei in the solar system (of the order of 10\% of the
corresponding elemental abundances), for which no satisfactory
explanation has been found so far (see \cite{argo03} for an
updated discussion of this problem). The remaining less
significant differences can be attributed to problems with the
nuclear physics involved or with an inappropriate choice and/or a
bad description of the stellar site(s) assumed to be at the origin
of the solar system p-nuclei. Any reduction in the nuclear physics
uncertainties will therefore put better constraints on the
determination of the astrophysical sites to consider. As far as
the global p-nuclei production is concerned and considering the
very large number of nuclear reactions involved in the production
of even a single p-nucleus, it is difficult to pinpoint one
critical experimental information on which nuclear physicists
should focus. Rather, any measurement of a photodisintegration cross section 
on nuclei which are located as
close as possible to the p-process path and for energies in the
appropriate Gamow windows should help gaining more confidence in
the calculated rates, in particular in getting direct insight into
the E1-strength function near particle threshold energies and in
obtaining more reliable astrophysical photodisintegration rates
without having to resort to detailed balance calculations.

Other problems raised by the p-process nucleosynthesis concern the
production of the rare odd-odd nuclei $^{180}$Ta$^{\rm m}$ and
$^{138}$La. In both cases \rgn\ reaction rates on these nuclei and
on their much more abundant neighbors $^{181}$Ta and $^{139}$La
are the essential nuclear quantities which will determine their
final abundances.

The rarest stable nucleus in the solar system and the only
naturally occurring isomer, $^{180}$Ta$^{\rm m}$, has been shown
by \cite{ray95} to be a natural product of the p-process in SNe-II. 
This conclusion was also largely shared
by \cite{rau02}, with different stellar models and updated
reaction rates. As seen in Sect.~\ref{sec:LCS}, the measurement of
the $^{181}$Ta($\gamma$,n)$^{180}$Ta reaction cross section at
energies close to the neutron threshold has provided a unique
opportunity to improve the description of the E1-strength
function and to obtain a more precise estimate of the
astrophysical rate for that reaction. The problem of the
$^{180}$Ta production has been re-examined in \cite{uts03} in the
case of a 25 M$_{\odot}$ model star with solar metallicity, using
the $^{181}$Ta($\gamma$,n)$^{180}$Ta rate constrained by the AIST
experiment. The previous prediction that $^{180}$Ta$^{\rm m}$ is
produced at the same level as the bulk of p-nuclides in SNe-II has
been quantitatively confirmed for a 25 M$_{\odot}$ SN-II and there are no 
reasons why different conclusions would be reached when the 
$^{180}$Ta$^{\rm m}$ productions calculated for other stellar masses will 
be averaged over a stellar mass function, as done in \cite{ray95}.  But on 
the other hand, the p-process origin of $^{180}$Ta$^{\rm m}$ has to be 
confronted to the fact that this
nuclide might \cite{gal98} or might not \cite{gormo00} be produced
by the s-process in AGB stars and that it is also expected to
receive some contribution from $\nu_e$-captures on pre-existing
$^{180}$Hf \cite{woos90}.

The rare odd-odd nucleus $^{138}$La is generally underproduced in
p-process calculations although it has been found recently that
exploding sub-Chandrasekhar-mass CO white dwarfs could be
significant $^{138}$La producers \cite{argo03}. The problem of the
$^{138}$La underproduction in more conventional p-process sites
like SNe-II has been addressed in \cite{gabr2001}. Two
reactions are critical for the thermonuclear production of that
nuclide, (1) $^{139}$La($\gamma$,n)$^{138}$La and (2)
$^{138}$La($\gamma$,n)$^{137}$La. In order to produce $^{138}$La
at the mean level of p-nuclide production in the 25 M$_{\odot}$
model star considered in \cite{gabr2001}, the ratio of the rates 
for reaction (1) to reaction (2) had to be increased by a factor 
20--25 with respect to the ratio obtained with HF (MOST) calculated 
rates. Such a large increase was very unlikely in view of an analysis 
of the nuclear physics uncertainties in the HF calculations.
On the other hand \cite{gabr2001} also re-examined the neutrino
production of $^{138}$La, originally proposed by \cite{woos90}, 
using an improved treatment of the (anti-)neutrino interactions, and
confirmed that neutrino processes can compensate the thermonuclear 
underproduction of $^{138}$La. However, if many of the input data 
necessary to calculate HF reaction rates have been measured for 
$^{139}$La, in contrast, very little is known experimentally for 
$^{137}$La or $^{138}$La. Clearly the measurement of \rgn\ cross 
sections on $^{139}$La and $^{138}$La are very desirable to 
disentangle the weak interaction and thermonuclear origins of $^{138}$La. 
Experimental values for the \rgn\ reaction on  $^{139}$La will soon be
available from recent measurements at AIST. Similar measurements
on $^{138}$La are a very stimulating challenge for the future!

\section{Perspectives}
\label{sec:outlook}

 Synchrotron radiation facilities of the third generation
are constructed in Europe (ESRF), America (APS) and Japan
(SPring-8). They feature a variety of insertion devices as light
sources.  At SPring-8, a 10 Tesla super-conducting wiggler (SCW)
was installed at the 8 GeV storage ring for a test production of a
high-energy radiation.

\begin{figure}[htb]
\begin{center}
\includegraphics [width=18pc]{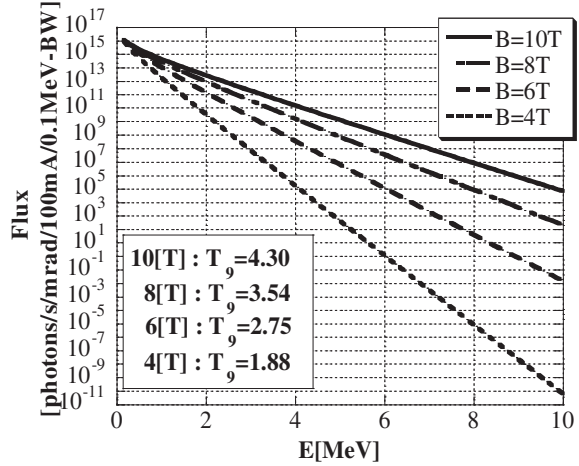}
\end{center}
\caption{ Synchrotron radiation from a 10 Tesla super-conducting
wiggler at SPring-8 \cite{cluster}. The temperature of the
black-body radiation equivalent to the high-energy part of the SCW
radiation is given for different magnetic fields. }
\label{fig:SCW}
\end{figure}

As shown in Fig.~\ref{fig:SCW}, this radiation
resulting from a 100 mA electron current is intense, even near
neutron thresholds around 8 MeV (10$^{7 - 8}$ photons sec$^{-1}$\, MeV$^{-1}$ for
a 10 T magnetic field) \cite{cluster}. More importantly, it is characterized by
exponential tails which mimic the high energy part of Planck
spectra corresponding to temperatures reached during the
p-process. The SCW radiation can thus be used to directly
determine the photonuclear reaction rate in Eq.~\ref{eq:rate} by
activation techniques without such manipulation as the
superposition of several bremsstrahlung spectra with different
end-point energies \cite{kv63}.

Alternatively, the experimental parameters ($\sigma_0$ and $\ell$)
involved in the threshold behavior of the photoneutron cross
section in Eq.~\ref{eq:thres} can be determined from a few
measurements of the reaction rate with the SCW radiation at the
highest available magnetic fields. Note that $\ell$ should be
treated as an experimental parameter, because of a possible mixture of
s- and p-wave neutron emissions.

There is a long list of photoreactions of interest for the
p-process (not listed here) which can be studied with the SCW
radiation at SPring-8. The photodisintegration of $^{180}$Ta is
certainly among them with a high priority.

The study of the photodisintegration of neutron deficient
radioactive nuclei along the p-process path is also one important
future project.  Coulomb dissociation into the neutron channel
where outgoing nucleus and neutron are measured in coincidence, is
a promising experimental technique, the development of which is
being considered at GSI. 

The bremsstrahlung and the monochromatic LCS beam play
complementary roles in the study of photon-induced reactions. With
its intense photon source, the former technique allows
photoactivation with natural target material because of the high
sensitivity to radioactive species.  A ground state photoreaction
rate is determined with a superposition method which approximates
the Planck distribution. The SCW radiation has the further
advantage to be also able to determine the threshold behavior of
\rgn\ cross sections. The latter photon source allows the cross
section measurement over the Gamow window, which is the most
direct way of testing the HF calculations of the rates. The direct
neutron counting can be applied to any nuclei in principle, but in
practice is limited to nuclei for which a considerable amount of
enriched target material can be made available. It must be
stressed however that there are presently certain regions in the
valley of stability that are accessible neither by
photoactivation nor by direct neutron counting, because some
photoreactions result in the production of stable nuclei or of
radioactive nuclei with extremely long half-lives, and because
target preparation is made difficult for nuclei with very small
natural abundances. Obviously, the emergence of a high-intensity
monochromatic photon source is awaited with great interest.
\section{Conclusions}
\label{sec:concl}

 Photon-induced reactions, ($\gamma$,x)
(x= n, p, $\alpha$, $\gamma$), have a direct impact on the
nucleosynthesis of the p-nuclei.  Among those, only ($\gamma$,n)
reactions of direct interest for astrophysics have been
investigated for selected nuclei, by photoactivation with
bremsstrahlung and by direct neutron counting with the LCS
$\gamma$-ray beam.  The experimental data enhance the reliability
and the predictive power of the Hauser-Feshbach model calculations
of the astrophysical reaction rates. The measurement of \rgn\ 
reactions on many more nuclei will follow, but the investigation
of ($\gamma$,$\alpha$) and ($\gamma$,p) reactions in the energy
range of interest is still a challenging prospect. In addition,
the E1 and M1 $\gamma$ strength functions below
particle thresholds should be addressed in direct relation to 
photoreactions on nuclear excited states under
stellar conditions.  Photoreactions on unstable nuclei are however 
beyond the present scope except with the virtual photon source, Coulomb
excitation/dissociation.

As demonstrated by the emergence of bremsstrahlung and laser
inverse Compton $\gamma$ rays, followed by the SCW synchrotron
radiation, the development of lasers, accelerators and
of related technologies will give fresh impetus to the creation of
new $\gamma$-ray sources of great value for nuclear astrophysics.

\begin{ack}
We want to thank H. Yonehara, J. Enders, U. Kneissl, A. Richter, T. Rauscher, and
S. Goriely for valuable discussions and private communications. 
The realization of the experiments at the S-DALINAC would not have been
possible without the enthusiasm of the members of the photon group, 
especially M. Babilon, D. Galaviz, K. Sonnabend, K. Volz, and K. Vogt.
The experiments at the AIST were made possible by the great efforts of H. Toyokawa, H. Ohgaki, K.Y. Hara, and S. Goko.
This work was supported by the Japan Society of the Promotion of Science, 
the Japan Private School Promotion Foundation, and the DFG under contract 
SFB 634.  M.R. is Research Associate of the National Fund for Scientific 
Research (Belgium).
\end{ack}

\end{document}